\documentclass[
    ,final            
  ]
  {aipproc}

\layoutstyle{6x9}

\begin{document}

\title{Phases of Dense Quark Matter and the Structure of Compact Objects}


\author{Andrew W. Steiner}{
  address={Theoretical Division, Los Alamos National Laboratory, Los
Alamos, NM 87545, USA}
}

\begin{abstract}
The presence of quark matter in neutron stars may affect several
neutron star observables and the neutrino signal in core-collapse
supernovae. These observables are sensitive to the phase of quark
matter that is present in compact objects. We present the first
calculation of the phase structure of dense quark matter which
includes a six-fermion color-superconducting interaction and show that
the effect of this term can destabilize the pairing interaction,
favoring phases where fewer quarks are paired. In turn, this
modification of the phase structure can modify the neutrino signal,
the structure of the neutron star, and the long-term cooling. We also
show that, contrary to the 20-year old paradigm of the surface
structure of the ``strange-quark stars'', the surface of these objects
may consist of nuggets of strange quark matter screened by the
electron gas.
\end{abstract}

\maketitle

The region of the QCD phase diagram at densities above a few times
$10^{14}$ g/cm$^3$ and temperatures on the order of 10 MeV, is highly
uncertain. Observations of neutron stars (or strange quark stars)
provide us with most of the constraints on the equation of state of
matter at these densities and temperatures. One of the most dramatic
features of QCD, the deconfinement phase transition, is not well
constrained by observations of compact objects. We may entertain a
couple possibilities:
\begin{enumerate}
\item All neutron stars may consist entirely of hadronic matter,
i.e. the central density of the maximum mass neutron star is below the
critical density for the deconfinement phase transition.
\item Some neutron stars may be sufficiently dense to contain deconfined 
quark matter in their interiors. Such neutron stars are often referred
to as ``hybrid quark stars''. Their central density is above the
deconfinement phase transition. 
\item The ``strange quark matter'' hypothesis may hold. Some (if not
all) neutron stars are, in fact, strange quark stars.
\end{enumerate}

I report on some progress in the theoretical description of hybrid
quark stars and strange quark stars and the associated observable
implications.


The description of dense quark matter has been recently revolutionized
by the observation that the color-superconducting gaps may be on the
order of 100 MeV~\cite{Alford98,Rapp98}. The essential features of all
models of gapped quark matter are the same. When the quark chemical
potential is much larger than the strange quark mass, then all nine
quarks participate in the pairing, giving the color-flavor-locked (or
CFL) phase. For smaller quark chemical potentials, the strange quark
mass is more important and only 4/9 color-flavor
combinations are paired. This is the two-flavor superconducting (or
2SC) phase.

It is known that a Lagrangian with four-fermion interactions alone
can not obey the same symmetries as QCD because the axial U(1)
symmetry is not broken by any four-fermion interaction. For this
reason, a six-fermion interaction, the {'}t Hooft interaction, is often
included
\begin{equation}
(\mathrm{a}) \qquad K~\mathrm{det}_f
\bar{\psi}_i \psi_j \bar{\psi}_k \psi_{\ell} \bar{\psi}_m \psi_n
\end{equation}
where $\mathrm{det}_f$ indicates a determinant over flavor space and
$K$ is a coupling constant. Although there is no reason to do so,
many authors have neglected a term of the form
\begin{equation}
(\mathrm{b}) \qquad K_{\mathrm{DIQ}}~\mathrm{det}_f
\bar{\psi}_i \psi^C_j \bar{\psi}^C_k \psi_{\ell} \bar{\psi}_m \psi_n
\, ,
\end{equation}
where $K_{\mathrm{DIQ}}$ is another coupling constant. This term is of
the same order in the fermion fields and also respects the symmetries
of QCD. Ref.~\cite{Steiner05} makes the first study of this term which
includes its effect on the quark masses and the gaps.

The basic conclusion is that for sufficiently large values of the
coupling constant, this color-superconducting 't Hooft interaction can
destabilize the color-superconducting gap. The effect is summarized in
Figure 1, where the masses and gaps are given as a function of the
coupling constant for a fixed density and temperature. The dramatic
behavior of these quantities at $K_{\mathrm{DIQ}}/K \sim 0.4$ causes
the gapped phases to become unstable. 

\begin{figure}[h]
  \includegraphics[height=8cm]{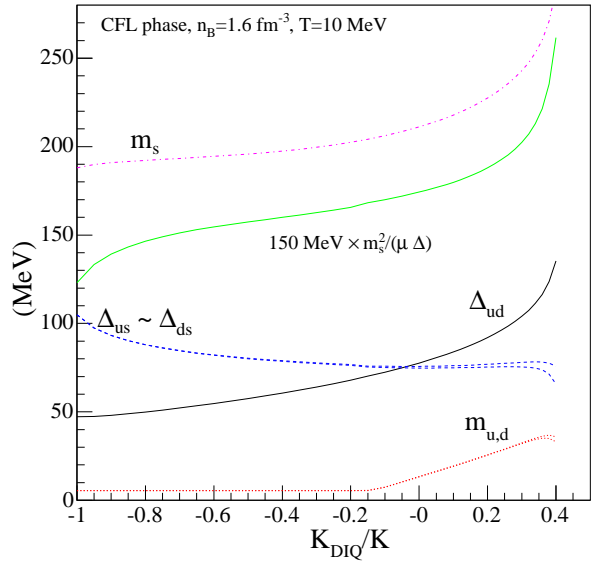}
  \caption{Quark masses and gaps versus $K_{\mathrm{DIQ}}/K$}.
\end{figure}

The presence (or lack) of a gap can affect the neutrino signal of a
newly-born proto-neutron star, the cooling properties of the neutron
star, and {\it any} observable which depends (or might depend) on the
magnitude of the size of the quark gaps in hybrid quark stars or
strange quark stars.


If the strange-quark matter hypothesis~\cite{Bodmer71,Witten84} is
correct, then strange quark matter may be the true ground state of
matter and nuclei are a metastable state. Neutron stars may be dense
enough to create a small amount of strange quark matter in the center
and the entire neutron star would then be converted from normal
hadronic matter to the strange quark matter ground state as neutrons
are absorbed by the strange quark matter in the interior. 

\begin{figure}[h]
  \includegraphics[height=7cm]{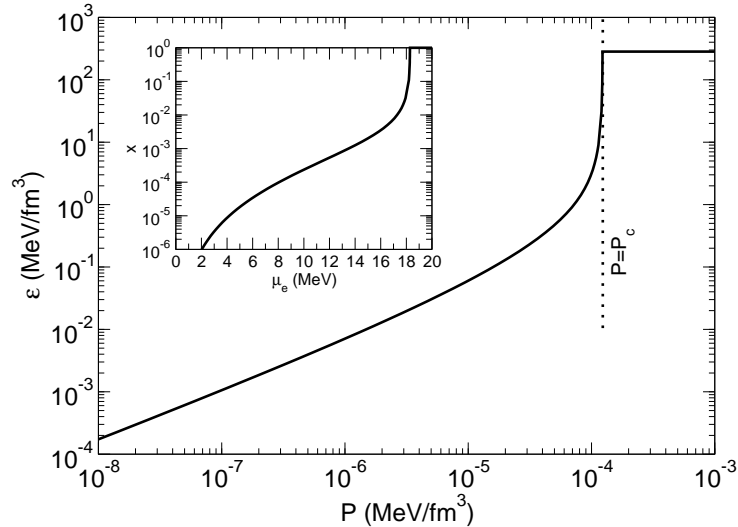}
  \caption{Equation of state}
\end{figure}

The long-standing paradigm for the description of strange quark stars
is that, because the pressure of strange quark matter vanishes at a
finite density, the surface of the strange quark star has a large
density discontinuity and a large electric
field~\cite{Page02}. However, if the surface tension between strange
quark matter and the vacuum is sufficiently small, this paradigm may
be incorrect~\cite{Jaikumar05} (also briefly mentioned in
Ref.~\cite{Haensel86}). In fact, small droplets (or ``nuggets'') of
strange quark matter may reside on the surface of a strange quark
star. These droplets create a more gradual drop in density over a
length of several meters above the neutron star. The corresponding
gradual drop in the energy density is displayed in Figure 2. The
energy density drops slowly at pressures below 10$^{-4}$ MeV/fm$^3$ in
contrast to the traditional picture where the energy density would
discontinuously vanish at $P=P_C$. Also plotted is the volume fraction
of strange quark matter in droplets as a function of the electron
chemical potential in the crust. The presence of the crust decreases
the electric field substantially and also decreases the photon
luminosity at the surface. Neutrino scattering off quark nuggets in
the crust will also modify the neutrino signal in supernovae.


I would like to thank Prashanth Jaikumar and Sanjay Reddy for many
helpful discussions. This work was supported by the Dept. of Energy under
contract W-7405-ENG-36.

\bibliographystyle{aipproc}   

\bibliography{panic}

\end{document}